# Trends in storm surge probability of occurrence along the Polish Baltic Sea coast

DOMINIK PAPROTNY

Delft University of Technology
Faculty of Civil Engineering and Geosciences
Stevinweg 1, 2628 CN, Delft, the Netherlands

e-mail: d.paprotny@tudelft.nl

**Abstract**

When assessing the hazard brought by storm surges on the coast, a frequency analysis is often conducted. An issue of particular concern is the sea level rise (SLR), thought to have an impact on maximum water levels. In this study, three gauge stations with long records dating back to the 19$^{th}$ century and located along the Polish Baltic Sea coast were analyzed. Different theoretical distributions were tested in order to find the best-fitting one. Gumbel distribution was chosen and estimated for 40-year moving periods. The location parameter soared in all tide gauges, but the shape parameter was stable in Gdańsk and decreased in the other two. In effect, theoretical annual maximum water levels followed the SLR in Gdańsk, whereas most extreme storm surges had no trend in Świnoujście and a negative one in Kołobrzeg. A possible explanation for this was investigated (change in atmospheric circulation measured by NAO index).

## 1. Introduction

In recent years researchers have undertaken increased efforts to identify the factors that control coastal morphology regardless of the type of coast. Numerous studies available (e. g. Ferreira *et al.*, 2009, Wziątek *et al.*, 2011, Lentz *et al.*, 2013) note the important role of extreme sea levels appearing during storm surges. In order to assess this source of hazard, a frequency analysis is commonly applied. It is used to calculate extreme sea levels with a certain probability of occurrence or, conversely, how often (on average) one can expect a particular storm surge to appear. Input data in this method are annual maximum water levels (AMWLs), which are then fitted to various statistical distributions in order to estimate flood likelihoods (FEMA, 2004). A 1% probability, known as the 100-year flood, is among the most commonly used in research and hazard mapping. In Poland, official maps of flood risk at the coast present a value of 0.2%, or 500-year return period (*Water Law Act*, 2001).

Alas, this method has a major drawback: it assumes that the investigated process is stationary, i.e. does not have a statistically significant trend (Ozga-Zielińska *et al.*, 1999). Meanwhile, current research show that mean sea levels are rising, both in the south Baltic Sea and in the world ocean. Wiśniewski *et al.* (2011) found that mean water levels at Polish tide gauges are increasing by 1–2 mm per annum, a similar pace to the global average (Church and White, 2011). This increase is commonly referred to as 'sea level rise' (SLR) and is believed to be triggered by the warming global climate (IPCC, 2013).

AMWLs are also changing over time. Data from Polish gauges with long time series (Kołobrzeg, Gdańsk and Świnoujście) are used for probability calculations (Wiśniewski and Wolski, 2009), even though they are hardly stationary. Student's *t*-test on Spearman's rank correlation coefficient disqualifies all those data from use in frequency analysis. Increasing AMWLs are attributed to higher mean sea levels, therefore some authors eliminate the SLR from the annual maxima series, creating "trend-eliminated" AMWLs. Wróblewski (1992, 1994) has done it for the Polish coast, also making forecasts based on the assumption that water levels with a certain return period will increase in the future by the same amount as annual mean sea levels. At the same time he indicated that this supposition is uncertain, as it implies that the statistical distribution of AMWLs does not change its shape over time. Notwithstanding, Xu and Huang (2013) recently used the same approach for stations in Florida.

The objective of this paper is to assess whether there is such a simple correlation between SLR and storm surges with a certain probability of occurrence along the Polish Baltic Sea coast using aforementioned tide gauges with more than 100 years of records. Given that calculations for stations with much shorter records are made, it would be therefore possible to divide the data into smaller subseries and conduct a frequency analysis in overlapping, moving intervals.

## 2. Material and methods

The three gauge stations used in this study were originally set up by Germany in the 19th century and are now under the authority of Poland (Figure 1). Data collected for the study are summarized in Table 1. Kołobrzeg has the longest AMWL series, though also the smallest amount of mean sea level data. Meanwhile tide gauge measurements from Świnoujście are one of the biggest mean water level datasets in the world, while the AMWL series is the shortest one in this study. Finally, Gdańsk has the longest concurrent series.

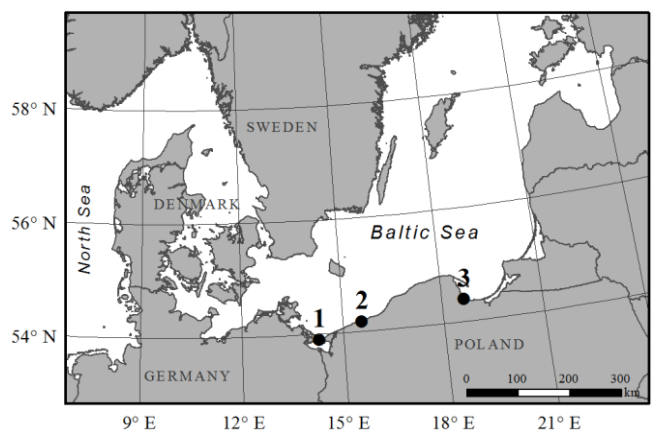

**Figure 1.** Location of tide gauges used in the study: Świnoujście – 1, Kołobrzeg – 2, Gdańsk – 3 (based on ERSI Maps)

**Table 1.** Dataset of annual maximum and mean water levels (years available).

| No. | Station | Annual maximum water levels | Annual mean water levels |
| --- | --- | --- | --- |
| 1 | Świnoujście | 1901–44, 1947–2007 | 1811–21, 1824–1944, 1947–2006 |
| 2 | Kołobrzeg | 1867–1943, 1946–2007 | 1901–40, 1946–2006 |
| 3 | Gdańsk | 1886–1939, 1946–2007 | 1886–1939, 1946–2006 |

AMWLs were extracted from Wiśniewski and Wolski (2009), while mean levels were taken from the *Permanent Service for Mean Sea Level* website, amended with information from Wiśniewski *et al.* (2011). Three years missing in mean sea level series for Kołobrzeg (1941–1943) were added using a correlation with data for Świnoujście in order to have a dataset consistent with the AMWL series (such approach was used, among others, by Wróblewski, 1974, Ekman, 2009 or Wiśniewski and Wolski, 2009).

It should be noted that the data refer to calendar years rather than water years. The stations' reference systems changed through the years, but Polish authors have already reduced all the data to a common datum (see Zeidler *et. al.*, 1995). They are presented here in reference to mean sea levels in Amsterdam, a measure known as the Normal Null (NN). Isostatic movements could possibly have an impact on the data, but Peltier (2004) and Hansen *et al.* (2012) show that the scale of the process at these particular locations is not significant. In fact, they could not alter the results of this study, because any vertical shifts would have modified both mean and maximum sea levels by the same factor.

As mentioned in the introduction, the entire dataset of AMWLs was divided into smaller subseries. A 40-year timespan was chosen, giving a total of 243 subseries (Świnoujście – 66, Kołobrzeg – 100 and Gdańsk – 77). 30-year periods are considered minimal for frequency analysis, whereas a minimum of 40 years used to be applied in Polish hydroengineering practice when calculating probabilities of less than 1% (Ozga-Zielińska *et al.*, 1999). Moreover, 40-year series gave better results in stationarity tests than longer time periods.

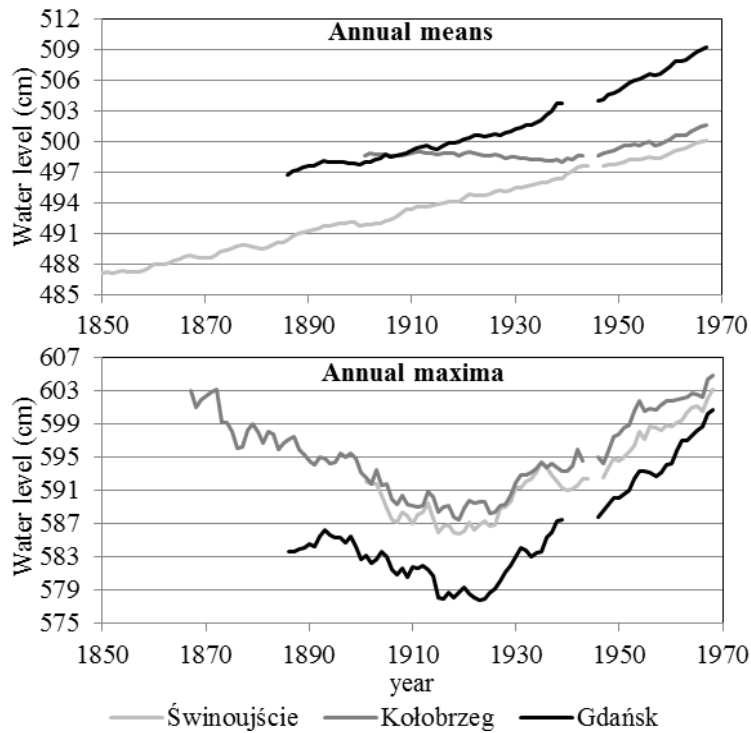

**Figure 2.** Trends in annual maximum and mean water levels, 40-year averages beginning with the year stated.

Baltic sea levels have periodical oscillations that could affect the data, but since their period ranges from 3 to 14 years (Wiśniewski *et al.*, 2011, Hagen and Feistel, 2005), they should not have an impact on the results. Long-term oscillations have been suggested (including a cycle of 90–100 years), however their exact nature—due to the length of time series—is uncertain and therefore will not be taken into consideration.

Variations of maximum and mean water levels in 40-year periods are presented in Figure 2, while long-term trends calculated by the author are shown in Table 2. Annual means in Gdańsk and Świnoujście have been increasing by more than 1 mm per year, whereas in Kołobrzeg a rate of only 0.3 mm per year was recorded. At the same time AMWLs were decreasing until mid-20$^{th}$ century, but then started rising at a considerably faster pace than mean levels – above 2 mm per annum. Still, the two measures are fairly well correlated, especially at Gdańsk station, albeit much better when 19$^{th}$ century data centuries are nor taken into consideration.

**Table 2.** Trend in maximum and mean water levels and the coefficient of determination ($R^2$) between maxima and means. Data for 40-year time periods beginning with the year stated. TR – trend in mm per year, SE – standard error (mm), * – trend not statistically significant ($p > 0.05$), x – information not available.

| Time periods | Station | Annual maximum water levels | | Annual mean water levels | | $R^2$ |
|---|---|---|---|---|---|---|
| | | TR | SE | TR | SE | |
| 1867-1967 | Świnoujście | x | x | 1.14 | 0.01 | x |
| | Kołobrzeg | 0.19 | * | x | x | x |
| 1886-1967 | Gdańsk | 1.80 | 0.20 | 1.47 | 0.04 | 0.69 |
| 1901-1967 | Świnoujście | 2.12 | 0.15 | 1.19 | 0.01 | 0.70 |
| | Kołobrzeg | 2.29 | 0.15 | 0.29 | 0.04 | 0.61 |
| | Gdańsk | 2.94 | 0.19 | 1.67 | 0.04 | 0.88 |

To qualify for a frequency analysis, data should checked using statistical methods. First of all, AMWL data should be tested for consistency and outlying elements. Here, this step was skipped, because the series were already verified, amended or otherwise cross-checked by several authors before (see Wiśniewski and Wolski, 2009). The next step is to analyse if the data are independent, i.e. they can possibly have any statistical distribution. Wald-Wolfowitz runs test was applied here. In this test, the number of runs—continuous sequences of data higher or lower than the median—was confronted with critical values at a confidence level of $\alpha = 0.05$. All but one subseries were found independent.

Last but not least, Student's *t*-test was used to verify if the AMWL data are stationary. The *t* statistic can be calculated with the following equation (1):

$$t = r_s \sqrt{\frac{(N-2)}{(1-r_s^2)}} \qquad (1)$$

where *N* is the number of elements and $r_s$ is Spearman's rank correlation coefficient (Ozga-Zielińska *et al.*, 1999). The results were confronted with a critical value which, at a confidence level of $\alpha = 0.05$ and *N* - 2 degrees of freedom, amount to 2.024 (Zwillinger, 2003). In spite of a total of 34 subseries being found not stationary—mainly data from Gdańsk—a frequency analysis was subsequently conducted, because a vast majority of data (86%) did pass the test.

Several statistical distributions are used in hydrology and oceanography, hence they had to be compared using a goodness-of-fit test. Firstly, the Kolmogorov-Smirnov (K-S) test utilized in order to determine whether the measurements could be approximated by certain theoretical distributions.

Subsequently, the Akaike Information Criterion (AIC) was applied to choose a distribution that fitted best with the data.

Four distributions were chosen: gamma, log-normal, Gumbel and Weibull (Ozga-Zielińska *et al.*, 1999, Wiśniewski and Wolski, 2009). The last two are combined in the generalized extreme value (GEV) distribution, however the results of probability calculations changed considerably between each 40-year period, whereas other distributions proved far less volatile to a solitary AMWL data point change. Consequently, GEV was not used in this study, as it apparently overfitted the data.

Parameters for each distribution and time period were estimated using the maximum likelihood method. Appropriate calculation procedures can be found in Ozga-Zielińska *et al.* (1999), except for Gumbel distribution, which was applied from Winner (2011). After the parameters were obtained, the K-S test was performed, checking a null hypothesis that the observed frequency distributions fitted the theoretical ones. The following equation was used (2):

$$D_{max} = \sup_{x}|f(x) - f_N(x)| \qquad (2)$$

where $f(x)$ is the theoretical, and $f_N(x)$ is the empirical cumulative distribution function (Zwillinger, 2003). The results are shown in Table 3.

**Table 3.** Results of the K-S test for 40-year time periods.

| Category | Station | Frequency distribution | | | |
|---|---|---|---|---|---|
| | | Gamma | Log-normal | Gumbel | Weibull |
| Maximum K-S test statistic for any time period | Świnoujście | 0.176 | 0.174 | 0.134 | 0.210 |
| | Kołobrzeg | 0.187 | 0.182 | 0.136 | 0.203 |
| | Gdańsk | 0.167 | 0.165 | 0.138 | 0.223 |
| Number of time periods failing K-S test | Świnoujście | 0 | 0 | 0 | 0 |
| | Kołobrzeg | 0 | 0 | 0 | 0 |
| | Gdańsk | 0 | 0 | 0 | 6 |

K-S test's critical value at a confidence level of $\alpha = 0.05$ and 40 elements is 0.210 (Zwillinger, 2003). Only a few subseries for Gdańsk failed the test for Weibull distribution. Otherwise the null hypothesis was not rejected. The second test, AIC, which checks goodness-of-fit, was used afterwards. The appropriate equation is as follows (3):

$$AIC = 2k - 2\sum_{j=1}^{N} \ln f(x_j) \qquad (3)$$

where $k$ is the number parameters (two for each distribution) and $f(x_j)$ is the probability density function for $j$ element (Mutua, 1994). The results are shown in Table 4. Gumbel distribution had, on average, the lowest AIC value, hence it fitted the data best. Weibull fared poorly, with the other two distributions giving only slightly worse results than Gumbel. Log-normal was found most reliable for almost one-third of the subseries, mainly those beginning after the 1930s. Nevertheless, Gumbel distribution was chosen for further analysis as the most accurate approximation of measured water levels.

**Table 4.** Results of the Akaike Information Criterion (AIC) for 40-year time periods.

| Category | Station | Frequency distribution | | | |
|---|---|---|---|---|---|
| | | Gamma | Log-normal | Gumbel | Weibull |
| Average AIC for all time periods | Świnoujście | 361.2 | 360.9 | 358.7 | 371.7 |
| | Kołobrzeg | 378.1 | 377.8 | 376.0 | 388.2 |
| | Gdańsk | 360.9 | 360.7 | 359.9 | 370.9 |
| Number of time periods having the lowest AIC value | Świnoujście | 0 | 27 | 39 | 0 |
| | Kołobrzeg | 1 | 37 | 62 | 0 |
| | Gdańsk | 12 | 25 | 40 | 0 |

This distribution was originally developed by E. J. Gumbel (1958), who created it especially for river discharge probability calculations. It has a location parameter $\alpha$ and a scale parameter $\beta$. Gumbel's distribution probability density function is given as follows (Weisstein, 2013):

$$P(x) = \frac{1}{\beta} \exp\left[\frac{x-\alpha}{\beta} - \exp\left(\frac{x-\alpha}{\beta}\right)\right] \qquad (4)$$

where $x$ is (in this case) the value of water level. In order to calculate water levels with a certain probability of exceedence $x_p$ equation (4) had to be transformed into (5):

$$x_p = \alpha - \beta \ln[-\ln(1-p)] \qquad (5)$$

where $p$ is the desired probability (Onoszko, 1992). For the sake of illustration, an example comparison between observed and theoretical water levels is presented in Figure 3. Measured water levels were given a probability using Weibull's plotting position formula (6):

$$p = \frac{m}{N+1} \qquad (6)$$

where $m$ is the element's rank in non-ascending order (Gordon *et al.*, 2004).

As the parameters $\alpha$ and $\beta$ were already estimated in order to compare the distributions' goodness-of-fit, water levels with any given probability could be easily obtained for the entire dataset.

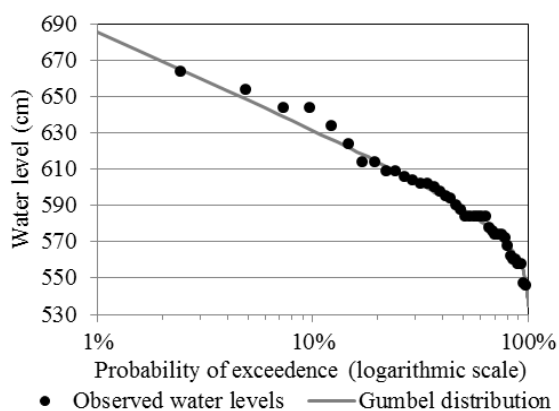

**Figure 3.** Empirical and theoretical (Gumbel distribution) AMWLs in Kołobrzeg, 1901–1940.

## 3. Results

Theoretical water levels with selected probabilities are presented in Figure 4. Additionally, linear trends were calculated and shown in Table 5. Even a brief look in the results reveals that the distribution of storm surges was changing over time. In Kołobrzeg water levels with a 500-, 100- or 10-year return period became lower since the 19th century. Only the most frequent storms (occurring every second year or more often) had an opposite tendency. In Świnoujście, extreme storms with the lowest probability of occurrence did not have a statistically significant trend, whereas all other were increasing with a pace proportional to its likelihood: 1 mm per annum for a 1% storm, 2 mm for a 20% one and almost 3 mm for a 99% one. Gdańsk is different than the other two, since water levels with different probabilities of exceedence were changing at a virtually identical speed of 2–2.2 mm per year. When only measurements made after 1901 are taken into account, the least probable storms present a more intense trend—up to 5.1 mm/year for a 0.1% storm—than the most probable ones (2.5 mm at 99% probability).

**Table 5.** Trends calculated for water levels with a certain probability of exceedence. Data for 40-year time periods beginning with the year stated. TR – trend in mm per year, SE – standard error (mm), * – trend not statistically significant (p > 0.05).

| Probability (%) | 1867–1968 Kołobrzeg | | 1886–1968 Gdańsk | | 1901–1968 Świnoujście | | 1901–1968 Kołobrzeg | | 1901–1968 Gdańsk | |
|---|---|---|---|---|---|---|---|---|---|---|
| | TR | SE | TR | SE | TR | SE | TR | SE | TR | SE |
| 0.1% | -4.7 | 0.3 | 2.2 | 0.5 | 0.3 | * | -2.1 | 0.4 | 5.1 | 0.5 |
| 0.2% | -4.1 | 0.3 | 2.2 | 0.5 | 0.5 | * | -1.6 | 0.4 | 4.8 | 0.5 |
| 0.5% | -3.4 | 0.3 | 2.1 | 0.4 | 0.8 | * | -0.9 | 0.4 | 4.6 | 0.4 |
| 1% | -2.8 | 0.2 | 2.1 | 0.4 | 1.0 | 0.5 | -0.4 | * | 4.4 | 0.4 |
| 2% | -2.3 | 0.2 | 2.1 | 0.4 | 1.2 | 0.4 | 0.1 | * | 4.1 | 0.3 |
| 5% | -1.5 | 0.2 | 2.1 | 0.3 | 1.5 | 0.3 | 0.8 | 0.2 | 3.9 | 0.3 |
| 10% | -0.9 | 0.2 | 2.1 | 0.3 | 1.7 | 0.3 | 1.3 | 0.2 | 3.6 | 0.3 |
| 20% | -0.3 | * | 2.0 | 0.2 | 2.0 | 0.2 | 1.8 | 0.2 | 3.4 | 0.2 |
| 50% | 0.6 | 0.2 | 2.0 | 0.2 | 2.3 | 0.1 | 2.6 | 0.1 | 3.1 | 0.2 |
| 75% | 1.2 | 0.1 | 2.0 | 0.2 | 2.5 | 0.1 | 3.1 | 0.1 | 2.8 | 0.2 |
| 99% | 2.1 | 0.1 | 2.0 | 0.1 | 2.9 | 0.1 | 4.0 | 0.1 | 2.5 | 0.1 |
| Location parameter α | 0.9 | 0.1 | 2.0 | 0.2 | 2.4 | 0.1 | 2.9 | 0.1 | 2.9 | 0.2 |

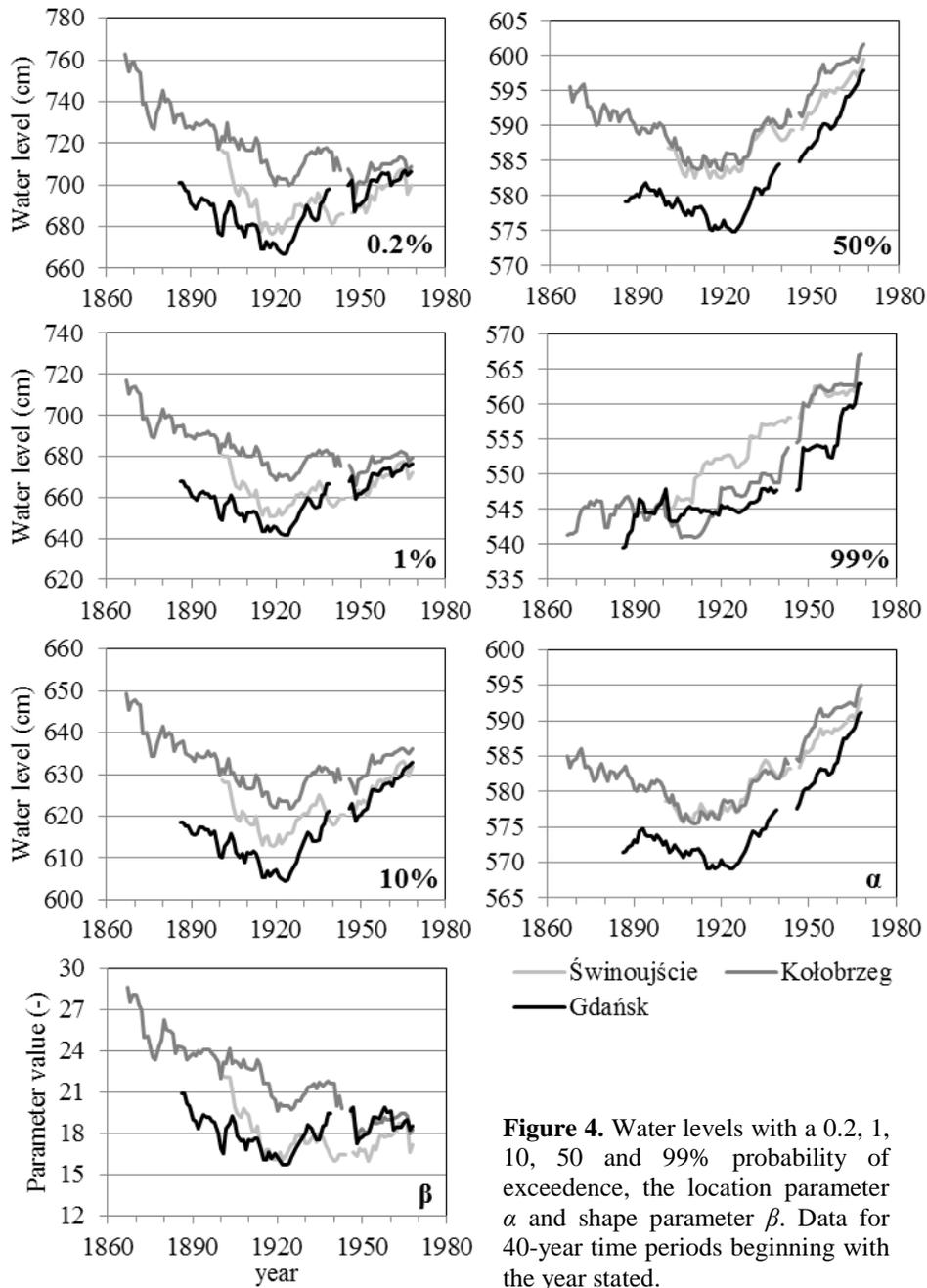

**Figure 4.** Water levels with a 0.2, 1, 10, 50 and 99% probability of exceedence, the location parameter $α$ and shape parameter $β$. Data for 40-year time periods beginning with the year stated.

The distinctive patterns of trends in Table 5 are caused by a shift in the distributions' parameters. The scale parameter $\beta$ decreased over time in Świnoujście and Kołobrzeg (Figure 4), which means that the distributions changed shape, becoming steeper and more compact. In Kołobrzeg it plummeted by more than a third, while in Świnoujście almost by a quarter. Data for Gdańsk indicate a different tendency: the scale parameter has barely changed since the 1886–1925 period and slightly increased since the early 20$^{th}$ century. Meanwhile the location parameter $\alpha$ soared in all tide gauges, most significantly of all in Gdańsk. The overall rate of change since the 1901–1940 period was 2.4 mm per year for Świnoujście and 2.9 mm for the other gauges. Data from earlier periods indicated an opposite trend, with a change of direction around the 1910s–1930s.

More interestingly, the relation between mean sea levels and aforementioned distributions of maximum water levels differs depending on the probability investigated. Figure 5 presents such a comparison, with the linear coefficient of determination shown in Table 6.

**Table 6.** Linear coefficient of determination between mean sea levels and water levels with a given probability of exceedence. Data for 40-year time periods beginning with the year stated. * – linear regression not statistically significant ($p > 0.05$).

| Probability (%) | $R^2$ | | | |
|---|---|---|---|---|
| | 1886–1967 | 1901–1967 | | |
| | Gdańsk | Świnoujście | Kołobrzeg | Gdańsk |
| 0.1% | 0.30 | 0.00* | 0.06 | 0.67 |
| 0.2% | 0.33 | 0.00* | 0.03* | 0.69 |
| 0.5% | 0.37 | 0.02* | 0.01* | 0.71 |
| 1% | 0.41 | 0.04* | 0.00* | 0.74 |
| 2% | 0.45 | 0.08 | 0.04* | 0.76 |
| 5% | 0.52 | 0.19 | 0.19 | 0.80 |
| 10% | 0.59 | 0.32 | 0.36 | 0.83 |
| 20% | 0.66 | 0.50 | 0.50 | 0.86 |
| 50% | 0.78 | 0.79 | 0.58 | 0.91 |
| 75% | 0.84 | 0.91 | 0.58 | 0.92 |
| 99% | 0.87 | 0.94 | 0.55 | 0.90 |
| Location parameter $\alpha$ | 0.81 | 0.86 | 0.58 | 0.92 |
| Scale parameter $\beta$ | 0.03 | 0.20 | 0.31 | 0.33 |

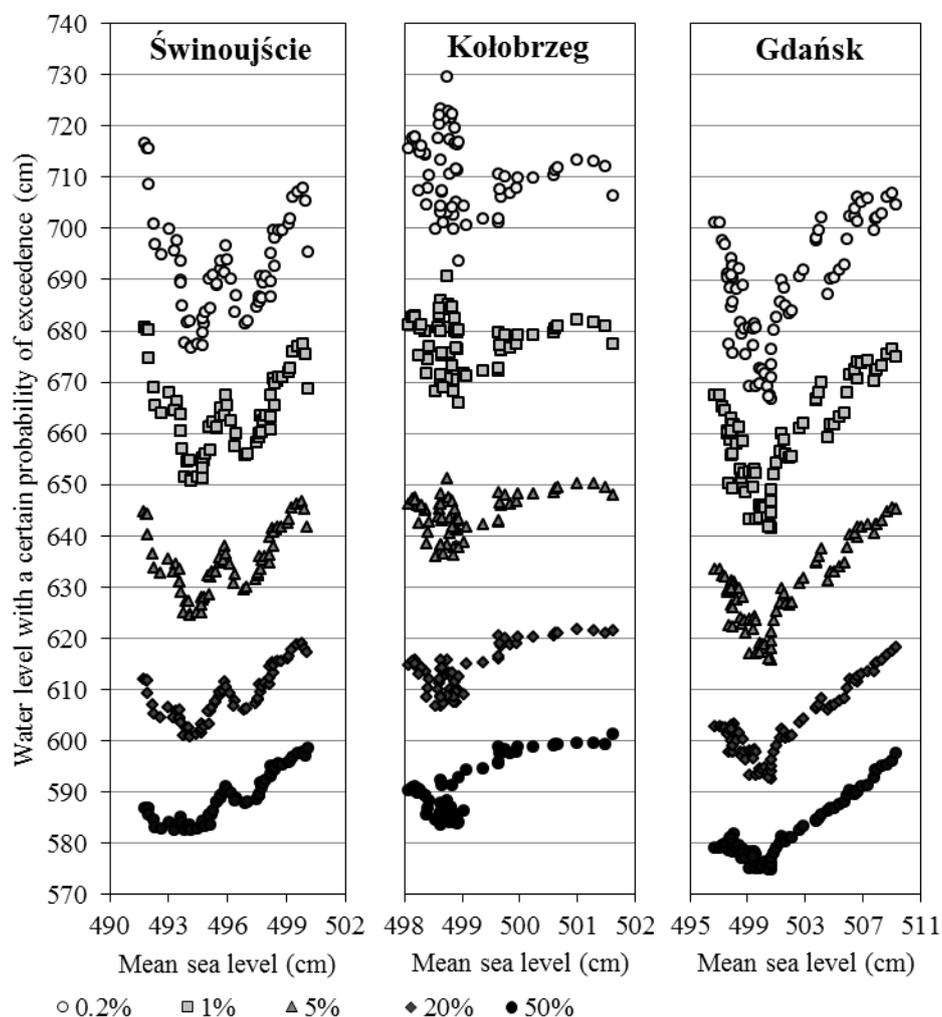

**Figure 5.** Comparison of mean water levels and water levels with a certain probability of exceedence. Data for 40-year time periods beginning with the year stated.

In most cases, $R^2$ is proportional to the probability of occurrence. In Świnoujście and Kołobrzeg there is virtually no correlation between mean sea levels and AMWLs with a return period of more than 30–50 years. In Świnoujście only storms that can possibly happen every second year or more frequently are well correlated with mean sea levels; in Kołobrzeg the relationship is even weaker. Meanwhile, data for Gdańsk show a much higher value of $R^2$ for extreme storms. If measurements made before 1901 are removed from the series, all mean sea levels in 40-year time periods have at least a fairly good correlation with theoretical AMWLs. The contrast between this and other stations in is striking: a 100-year storm has a $R^2$ value of 0.74 in Gdańsk, 0.04 in Świnoujście and near zero in Kołobrzeg.

The two parameters of Gumbel distribution show very different relationship: α (location) fits well with mean sea levels while β (scale) does not. In Świnoujście and Gdańsk the location parameter correlates strongly with mean water levels. Kołobrzeg, on the other hand, has a lower $R^2$ value, but it should be noted that the annual means series are themselves an oddity compared to the other two tide gauges (Figure 2), let alone the rest of south Baltic stations.

Generally, AMWL distributions covering the 19th and early 20th century are either uncorrelated (Kołobrzeg) or even negatively correlated (i.e. higher mean sea levels appear with lower theoretical AMWLs). On the other hand, when only measurements starting from ca. 1910 are included, the coefficient of determinations in Świnoujście and Gdańsk increases up to 0.92 and 0.97, respectively. Therefore, it can be concluded that there is no simple and obvious correlation between mean and extreme sea levels.

**Discussion and conclusions**

Storm surge parameters in the Baltic Sea are a result of several factors, such as wind direction and velocity, air pressure in baric lows and the volume of water in the Baltic. The latter is largely influenced by water inflows from the North Sea via the Danish straits, themselves caused by strong westerly winds (Ekman, 2009). The relationship between North Atlantic Oscillation (NAO) index, a method to measure atmospheric circulation in Europe, and sea levels in the Baltic is well-researched, most recently by Stramska and Chudziak (2013). They note that the correlation between mean sea levels and winter NAO index changes significantly between different locations in the basin.

Consequently, the correlation between Hurrell's (2013) NAO index and theoretical AMWLs has been calculated, as a possible explanation of different results in each station, as mentioned in the previous section. A maximum positive value of NAO from monthly means in a year was used here instead of a winter or annual average. In was found that the correlation with theoretical AMWLs is weak and decreasing eastwards, as can be seen in Table 7. In Świnoujście the coefficient of determination for a 0.1% (1000-year) storm is 0.29, compared to 0.17 in Kołobrzeg and a mere 0.02 in Gdańsk. However, the $R^2$ value for scale parameter β in Świnoujście is a moderate 0.46. In Kołobrzeg the correlation is weak (0.14) and ever smaller than that in Gdańsk (0.02). A fairly strong link between NAO and β might explain the disparity in correlation of the least and the most probable storm events in Świnoujście, as seen in Table 6. Annual maxima of NAO decreased slightly over time, so this change in atmospheric circulation could have offset the impact of sea level rise (SLR) on extreme storm surges.

**Table 7.** Linear coefficient of determination between Hurrell's NAO index and water levels with a given probability of exceedence. Data for 40-year time periods beginning with the year stated. * – linear regression not statistically significant ($p > 0.05$).

| Probability (%) | $R^2$ (1901–1968) | | |
|---|---|---|---|
| | Świnoujście | Kołobrzeg | Gdańsk |
| 0.1% | 0.29 | 0.17 | 0.02* |
| 0.2% | 0.26 | 0.17 | 0.02* |
| 0.5% | 0.23 | 0.16 | 0.02* |
| 1% | 0.19 | 0.14 | 0.02* |
| 2% | 0.15 | 0.11 | 0.02* |
| 5% | 0.08 | 0.05 | 0.01* |
| 10% | 0.03* | 0.01* | 0.01* |
| 20% | 0.00* | 0.00* | 0.01* |
| 50% | 0.03* | 0.02* | 0.01* |
| 75% | 0.09 | 0.03* | 0.01* |
| 99% | 0.22 | 0.06 | 0.00* |
| Location parameter $\alpha$ | 0.06 | 0.03* | 0.01* |
| Scale parameter $\beta$ | 0.46 | 0.14 | 0.02* |

Different results obtained for each station renders them variously subjectable to long-term forecasting. Gdańsk's tide gauge is closest to assumptions made by Wróblewski (1994) or Zeidler *et al.* (1995) that future theoretical AMWLs at the Polish coast will increase at the same pace as the SLR. It is because the location parameter $\alpha$ in this station is very well correlated with mean sea levels and the scale parameter $\beta$ has no statistically significant trend. Nevertheless, $\alpha$ has risen there by 1.36 mm per year for every 1 mm of mean sea level increase between the first and the last 40-year period. Given that the SLR in the 21st century is likely to be anywhere between 2.7 and 10.2 mm per year (IPCC, 2013), a 1% storm in 2081–2120 could be 714–837 cm high, compared to 670 cm in 1961–2000.

In Świnoujście and Kołobrzeg the scale parameter changed considerably since the beginning of the 20th century. Storms of ca. 2% or lower probability of occurrence have no trend in the former and a negative one in the latter. More frequent storms have become much higher. In light of evidence shown above, it can be concluded that typical storm surges have become higher due to SLR, while the most extreme ones did not because of less violent weather, as indicated by lower NAO maxima. In most recent decades NAO started rising, and so have storms with low probability of occurrence. Therefore, though the location parameter and high frequency storms in those two stations increase at a pace similar to the gauge in Gdańsk, forecasting would require also a prediction of change in atmospheric circulation.

In total, the three analysed tide gauges can be characterized as follows:

- Distribution of storms in Świnoujście is heavily influenced by both mean sea level (well correlated with the location parameter) and atmospheric circulation (fairly good correlation of NAO index with the scale parameter).
- Kołobrzeg station shows much lower (but still good) correlation between Gumbel distribution of AMWLs and mean water levels than the other two tide gauges and vaguely correlated with NAO.
- Gdańsk's theoretical AMWLs are in very close match with mean sea levels and completely unrelated to NAO index.

An overall conclusion for all stations is that:

- AMWLs with a high probability of occurrence are changing in close relation to SLR;
- AMWLs with a low probability of occurrence are related to atmospheric circulation, though this link is decreasing eastwards.

Nevertheless, aforementioned conclusions should be subject to further research, in the means of statistical analysis of other Baltic stations will long records.